\pdfoutput=1
\documentclass[11pt,a4paper]{article}

\usepackage[margin=0.1in,font=small]{caption}
\usepackage[affil-it]{authblk}
\usepackage[top=1in, bottom=1.25in, left=1.5in, right=1.5in]{geometry}
\usepackage{graphics}
\usepackage{graphicx}
\usepackage{pifont}
\usepackage{siunitx}
\usepackage{amssymb}
\usepackage{amsmath}
\DeclareSIUnit\gauss{G}

\title{\LARGE Difference-frequency combs in cold atom physics}
\author[1]{\large Russell Kliese}
\author[1]{\large Nazanin Hoghooghi}
\author[1]{\large Thomas Puppe}
\author[1]{\large Felix Rohde}
\author[1]{\large Alexander Sell}
\author[1]{\large Armin Zach} 
\author[1]{\large Patrick Leisching}
\author[1]{\large Wilhelm Kaenders}

\author[2]{\large Niamh C. Keegan} 
\author[2]{\large Alistair D. Bounds}
\author[2]{\large Elizabeth M. Bridge}
\author[1,2]{\large Jack Leonard} 
\author[2]{\large Charles S. Adams}
\author[2]{\large Simon L. Cornish} 
\author[2]{\large Matthew P. A. Jones}

\affil[1]{\footnotesize TOPTICA Photonics AG, Lochhamer Schlag 19, 82166 Graefelfing, Germany}
\affil[2]{\footnotesize Joint Quantum Centre (JQC) Durham-Newcastle, Department of Physics, Durham University, Durham DH1 3LE, UK}

\date{\small May 4, 2016}

\begin{document}

\maketitle

\begin{abstract}
Optical frequency combs provide the clockwork to relate optical frequencies to radio frequencies. Hence, combs allow to measure optical frequencies with respect to a radio frequency where the accuracy is limited only by the reference signal. In order to provide a stable link between the radio and optical frequencies, the two parameters of the frequency comb must be fixed: the carrier envelope offset frequency $f_{\rm ceo}$ and the pulse repetition-rate $f_{\rm rep}$. We have developed the first optical frequency comb based on difference frequency generation (DFG) that eliminates $f_{\rm ceo}$ by design - specifically tailored for applications in cold atom physics. An $f_{\rm ceo}$-free spectrum at \SI{1550}{\nano \meter} is generated from a super continuum spanning more than an optical octave. Established amplification and frequency conversion techniques based on reliable telecom fiber technology allow generation of multiple wavelength outputs. In this paper we discuss the frequency comb design, characterization, and optical frequency measurement of Sr Rydberg states. The DFG technique allows for a compact and robust, passively $f_{\rm ceo}$ stable frequency comb significantly improving reliability in practical applications.
\end{abstract} 

\section{Introduction}
\label{intro}
An exact measurement of a frequency requires a clockwork to establish a phase coherent link between the frequency to be measured and a reference oscillator of known frequency. An optical frequency comb is an elegant solution providing such a clockwork between two optical frequencies or an optical and a microwave (RF) frequency \cite{Cundiff:03,Diddams:10}. It is hence an essential building block for practical implementations of optical clocks \cite{Bartels:05,Ludlow:15,Poli:13}. Other applications include precision measurements, generation of arbitrary optical frequencies and the transfer of optical and radio frequencies. Quantum optics experiments require accurate measurement and stabilization of optical frequencies for preparation, precise control and probing of the quantum system. With cumulative sophistication of experimental schemes and required control, the light fields span increasingly large ranges in wavelength. For example in Rydberg experiments, in addition to the first excited states used for initial preparation, a number of higher excited levels are addressed. Note, that while ground state transitions can be accessed, e.g., with Doppler-free absorption spectroscopy, this is significantly more difficult for transitions to higher states. Using the comb all frequencies can be linked to a common reference oscillator. To reference different wavelengths multiple corresponding outputs of the comb are generated by appropriate frequency conversion. The DFG comb presented here was developed by TOPTICA in collaboration with the atomic and molecular physics groups at Durham University for applications in experiments on high-precision spectroscopy of Sr Rydberg states \cite{Bridge:16} and ultracold RbCs molecules \cite{Kumar:16}.  The laser wavelengths involved span from the UV to the NIR.

Generally both the center frequency as well as the phase noise or linewidth of the individual laser line are of interest. The short term stability, as characterized by the linewidth, can be reduced by a short term reference such as a high-finesse cavity. The long term stability, commonly characterized by the Allan variance, can be provided by a nearby spectroscopic feature, however this is not generally available and is sensitive to environmental parameters. A frequency comb on the other hand can be referenced to a GPS disciplined RF oscillator, providing absolute accuracy exceeding state-of-the-art commercially available wavelengthmeters by orders of magnitude. In this paper we review properties of frequency combs as a tool for quantum optics, and discuss in detail the DFG comb and an expample application.

\section{How the frequency comb works}
\label{sec:2}
\begin{figure}
\includegraphics[width=\textwidth]{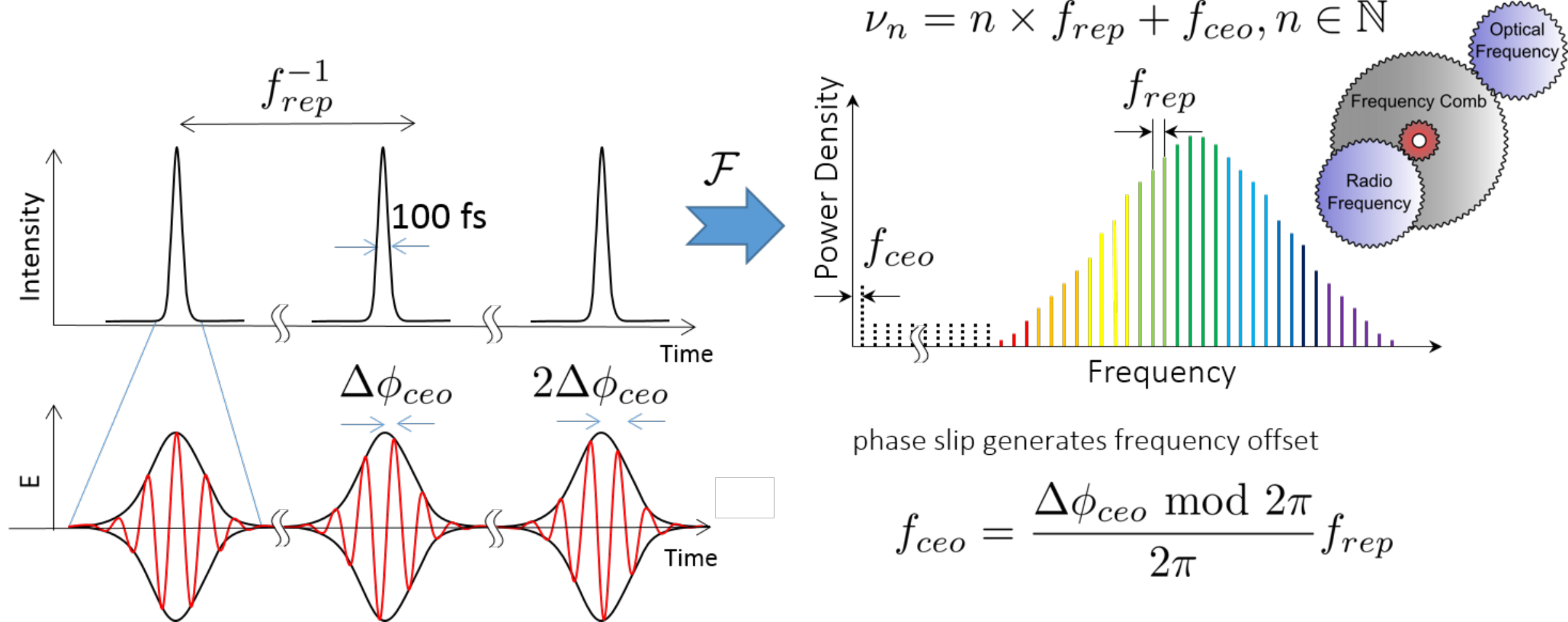} 
\caption[width=0.9\textwidth]{The optical frequency comb: The comb spectrum generated from a mode-locked short-pulse laser is completely defined by the repetition rate $f_{\rm rep}$ and the carrier envelope offset frequency $f_{\rm ceo}$. It provides a clockwork to link optical and RF frequencies.}
\label{fig:fc}       
\end{figure}

Optical frequency combs are typically based on passively mode-locked oscillators emitting a train of pulses separated by the round trip time of the optical resonator. The optical phases of the frequency components supported by the resonator are phase locked by the mode-lock mechanism to form short pulses. Short pulses are commenly generated using Ti:Sapphire oscillators ($<\!\SI{10}{\femto \second}$) \cite{Rausch:08,Xu:97}, as well as erbium doped fiber oscillators ($<\!\SI{100}{\femto \second}$) \cite{Sell:09,Tamura:93} and are fundamentally Fourier-limited by the gain bandwidth. In the frequency domain the pulse train results in a spectrum of equidistant modes separated by the repetition rate $f_{\rm rep}$ given by the inverse round-trip time (see Fig.\;\ref{fig:fc}). Due to the difference of phase and group velocities within the oscillator, the phase of the optical carrier shifts with respect to the pulse envelope from pulse-to-pulse. This gives rise to the carrier envelope phase shift which amounts to a non-zero frequency offset $f_{\rm ceo}$ in the Fourier spectrum. Hence the comb spectrum generated from a mode-locked short-pulse laser is completely defined by these two parameters. The frequency of the comb lines generated by a mode-locked pulsed laser $\nu_n$, are given by:
\begin{equation}
\nu_{n}=n \times f_{\rm rep}+f_{\rm ceo},
\label{eq:nun}
\end{equation}
with $n \in \mathbb{N} $ denoting the comb tooth order.

The repetition rate is directly linked to the optical path length in the resonator and hence an actuator for the cavity length is the obvious choice. The carrier envelope frequency $f_{\rm ceo}$, i.e. a pulse-to-pulse phase shift can be influenced by acting on group velocity dispersion in the oscillator internally, e.g. by pump current modulation \cite{Jones:00}. Alternatively $f_{\rm ceo}$ can be stabilized externally by a frequency actuator, e.g. an acousto-optical modulator (AOM) \cite{Koke:10}, which can also be implemented in a feed-forward scheme. Note that care has to be taken to avoid cross talk \cite{Dolgovskiy:12,Zhang:12} in both the detection as well as regarding the actuators. Since the relative phase is typically detected by conversion to amplitude, the detection is sensitive to amplitude to phase noise conversion. A well-established scheme to measure the carrier envelope offset frequency $f_{\rm ceo}$ is an $f\!\!-\!\!2f$ interferometer, where $f_{\rm ceo}$ is detected via a beat between the frequency-doubled low frequency end and the high frequency end of an octave spanning spectrum. 

In a difference-frequency comb, instead of detecting and stabilizing $f_{\rm ceo}$ in a feed-back loop, it is fundamentally removed by difference frequency generation (DFG), leaving only a single parameter $f_{\rm rep}$ to be stabilized. The performance of a frequency comb is given by the intrinsic noise properties of the oscillator and the ability to detect and stabilize $f_{\rm rep}$ and $f_{\rm ceo}$. The noise properties of the oscillator are mainly determined by the type and, in particular, the time scale of the mode-locking mechanism and the balance between gain and outcoupling. If the contributions of a noise source to $f_{\rm rep}$ and $f_{\rm ceo}$ are linearly correlated, the induced frequency fluctuation can be described by the elastic tape model (ETM) \cite{Telle:02}. In this case the comb modes show a breathing motion with respect to a fix point. The fix point is given by the cancellation of the contributions, i.e. at the tooth index $n_{\rm fix} = −(df_{\rm ceo}/dN)/(df_{\rm rep}/dN)$, where $N$ is the parameter varied. As pointed out in \cite{Benkler:05} noise sources non-compliant with the ETM could significantly complicate both the full stabilization of the comb by phase locking as well as accurate frequency measurements. For the type of Er:fiber oscillators similar to that described in \cite{Tamura:93,Sell:09} the observed fix-point for pump current modulation is at optical frequencies close to its emission spectrum \cite{Haverkamp:04}. Hence, in this case a piezo or fiber stretcher acting on the cavity length and the pump current amount to two nearly orthogonal actuators. In reality however the different noise sources and actuators will have different fix-points, moreover these fix-point will dependent on the modulation frequency and hence lead to multiple frequency dependent fix-points \cite{Dolgovskiy:12,Benkler:05}.    
 
\section{The difference frequency comb}
\label{sec:3}

DFG combs have previously been realized in pulsed Ti:Sapphire lasers to generate $f_{\rm ceo}$-free spectra in the IR \cite{Fuji:04,Zimmermann:04} as well as Ytterbium doped fiber combs. A technologically elegant solution \cite{Fehrenbacher:15,Krauss:11} is based on a more than octave spanning super continuum such that the DFG output is at the original erbium oscillator spectrum at \SI{1550}{\nano \meter}. This allows use of the same frequency conversion techniques, based on reliable telecommunications fiber components, previously developed for erbium fiber oscillators.

\begin{figure}
\includegraphics[width=\textwidth]{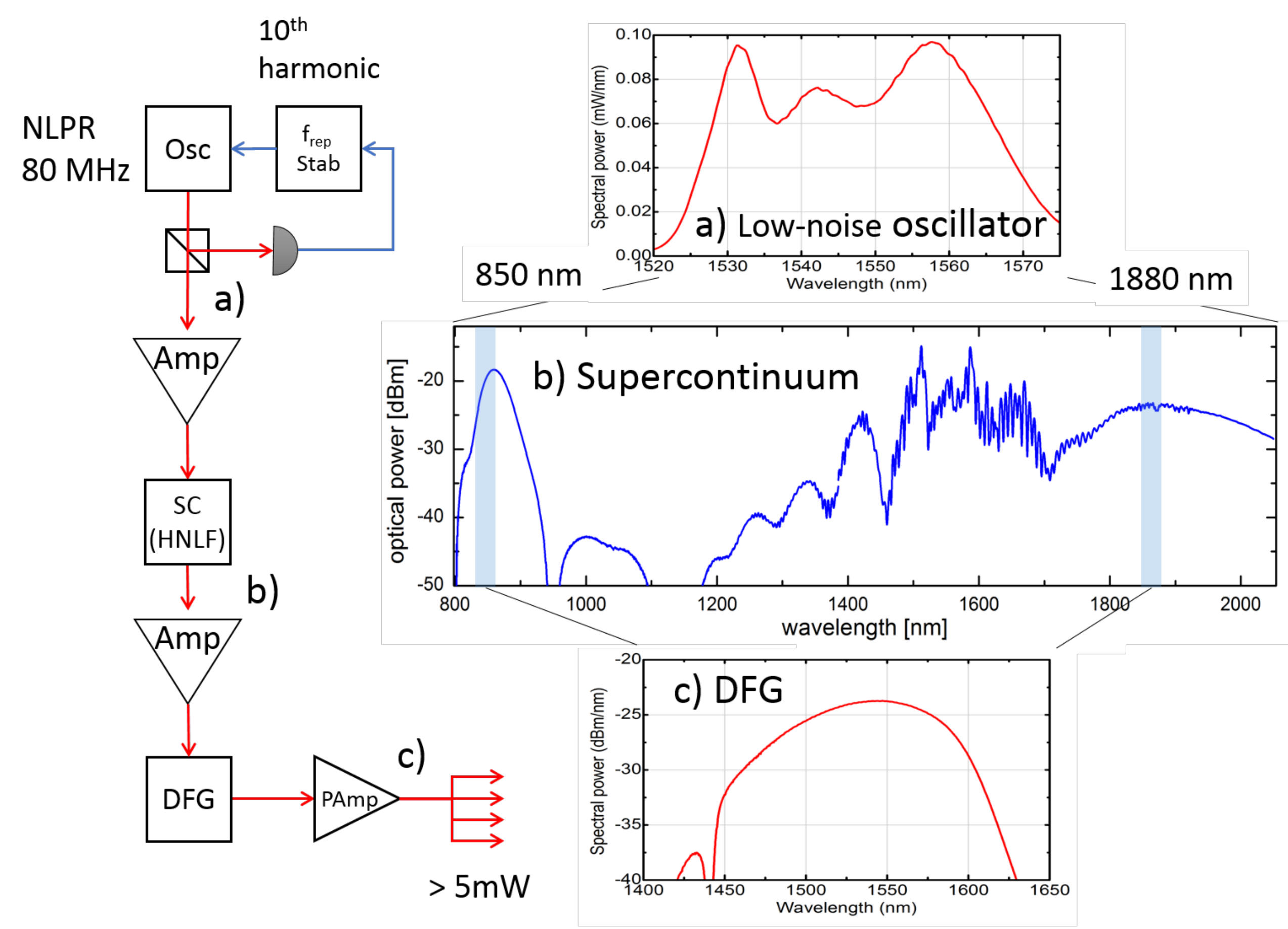}
\caption{DFG comb: In a difference frequency comb (DFG-FC) the fceo is removed by mixing two parts of the spectrum. In this particular DFG-FC super continuum spanning more than an optical octave (b) is generated from the low-noise oscillator (a) using a highly nonlinear fiber (HNLF). Difference freqeuncy generation (DFG) between the ends at \SI{850}{\nano \meter} and  \SI{1880}{\nano \meter} results in a passively fceo-stable output at \SI{1550}{\nano \meter} (c). The $f_{\rm ceo}$-free comb spectrum at \SI{1550}{\nano \meter} can then be amplified and frequency converted with established techniques based on Er:fiber technology.}
\label{fig:dfg}       
\end{figure}

Fig. \ref{fig:dfg} illustrates the process of generating the offset-free comb from the spectrum of a fiber oscillator. First the spectrum is amplified and broadened in a highly non-linear fiber (HNLF) to include \SI{850}{\nano \meter} and \SI{1880}{\nano \meter}. The difference frequency of these parts of the spectrum is subsequently generated in a periodically poled lithium niobate (PPLN) non-linear crystal. Details of the setup can be found in \cite{Fehrenbacher:15}. Since the comb teeth are coherent the difference of multiple combinations of lines $ (n-m)\equiv\bar{n}=const.$ contribute to a single line in the DFG spectrum according to:

\begin{eqnarray}
\underbrace{\nu_n- \nu_m}_{\substack{\nu_{\bar{n}}}} &=& \underbrace{(n-m)}_{\substack{\bar{n}}} \times f_{\rm rep}+\underbrace{(f_{\rm ceo}-f_{\rm ceo})}_{\substack{\equiv 0}} \\
  \nu_{\bar{n}}  &=& \bar{n} \times f_{\rm rep},
\label{eq:dfg}
\end{eqnarray}

After the DFG the $f_{\rm ceo}$ is equal to zero and the comb lines are integer multiples of $f_{\rm rep}$ (eq. \ref{eq:dfg}). By characterizing the noise properties of individual comb teeth it can be shown that the fix point of the DFG comb is to very good approximation at the frequency origin \cite{Puppe:16}. Hence, the $f_{\rm rep}$ noise measured at one frequency can be rescaled to give the properties at any comb line.

\section{Linking cw laser and frequency comb}
\label{sec:4}

There are two common ways to link a frequency comb with a cw laser. Either the frequency comb can be locked to a stable reference and in turn used as a reference for measuring an unknown frequency of a cw laser, or a cw laser can be locked to the comb to provide a precisely defined cw laser frequency.

 \begin{figure}
\includegraphics[width=\textwidth]{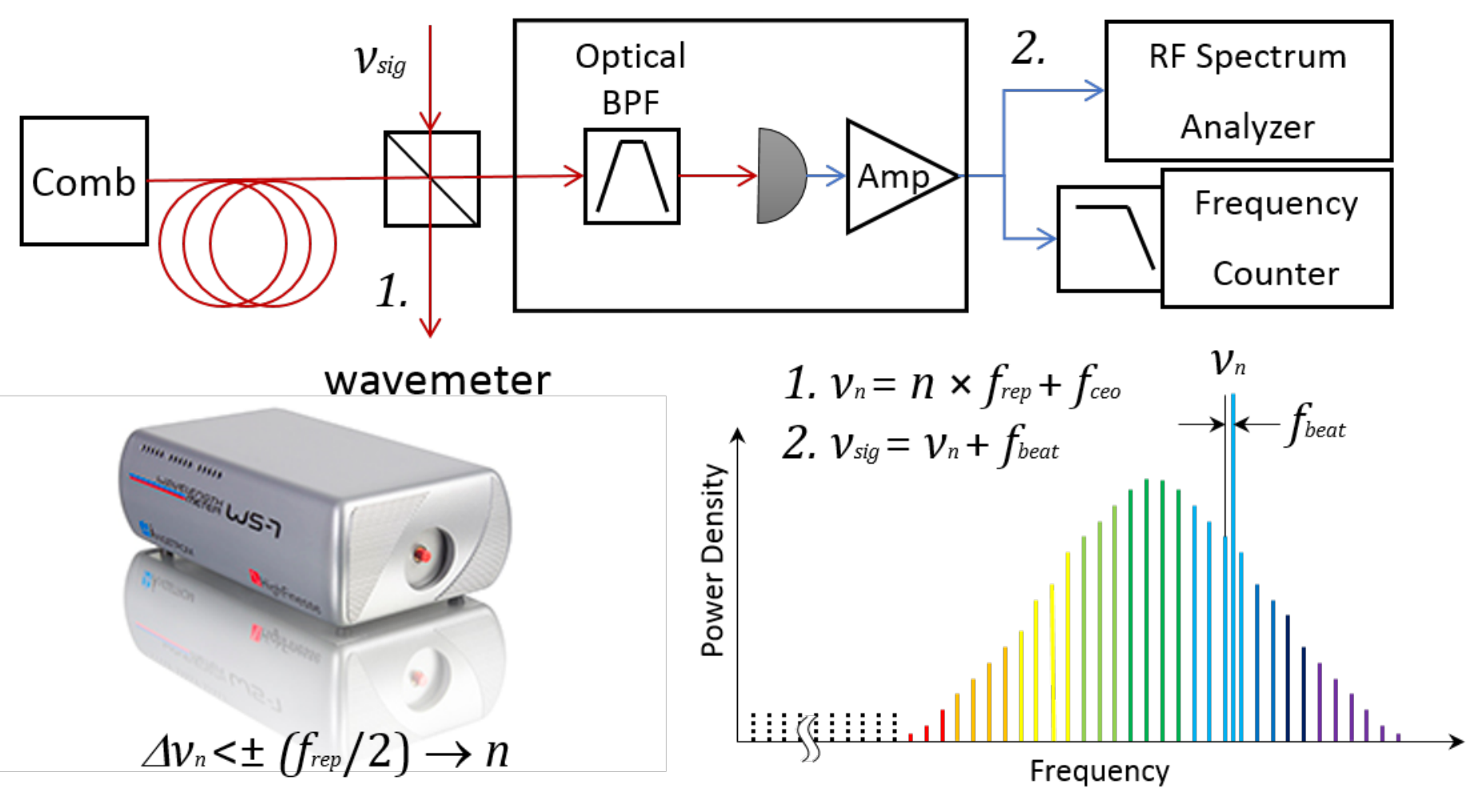}
\caption{Measurement of the frequency of a cw laser with a frequency comb typically includes two steps: 1. Measurement of the frequency to within one $f_{\rm rep}$ to identify the order of a nearby comb tooth, e.g., with a wavelengthmeter, 2. Measurement of the beat between the laser and this comb tooth as an RF offset frequency. For locking the laser to the comb the beat offset frequency is phase locked.}
\label{fig:beat}       
\end{figure}

To measure an unknown optical frequency of a cw laser with a frequency comb, the frequency comb is typically locked to an RF reference. A schematic of the components involved in locking the frequency comb to an RF reference is shown in (Fig.\,\ref{fig:setup}). The repetition rate is detected by a photodetector and a phase error signal is generated by mixing the repetition rate with a stable RF reference. The phase error in turn acts on the optical oscillator, for example by actuating a piezo actuator, forming a phase-locked loop (PLL) to stabilize the repetition rate of the frequency comb to the RF reference. In the DFG comb,  the fix point is at the frequency origin, hence locking the repetition rate to a  RF reference provides an absolutely stable frequency comb, typically limited only by the performance of the RF reference. The measurement of the unknown frequency of the cw laser is depicted in (Fig.\,\ref{fig:beat}) and consists of two steps.  First, the index, $n$, (eq.\,\ref{eq:dfg}) of a close-by comb line is determined using a wavelengthmeter, and then the beat frequency of the cw laser and the comb line is determined to provide the precise offset from the previously determined comb line index. To improve the SNR of the beat, the relevant part of the comb spectrum is isolated by an optical bandpass filter that reduces the optical power incident on the beat photodiode from the rest of the comb.

\begin{figure}
 \includegraphics[width=\textwidth]{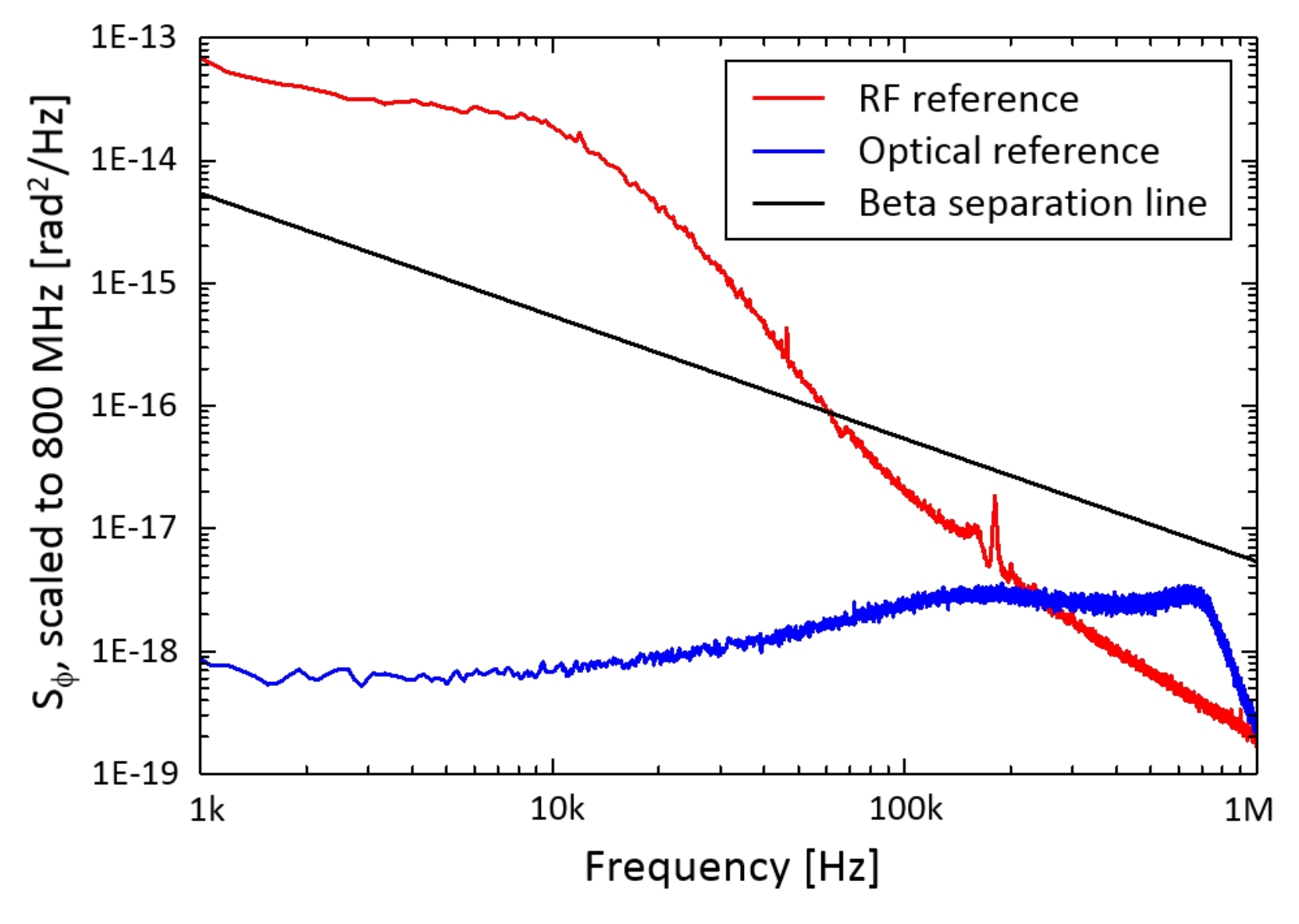}
\caption{Phase noise measurements of the DFG frequency comb at 1162 nm with the comb locked to an RF reference (red) and a narrow line-width optical reference (blue). When locked to the optical reference, a Hz level linewidth of the comb lines is achieved. The beat seperation line (black) is shown for reference.}
\label{fig:rf_opt}       
\end{figure}

The noise limitation due to a RF reference is significant at optical frequencies,  since there is a $10^6$ scaling to optical frequencies given by the respective comb order, $n$.  A much improved noise performance can be achieved by locking $f_{\rm rep}$ to an optical reference \cite{Bartels:03}. This greatly enhances the sensitivity of detection and improves the signal to noise ratio (SNR). The repetition rate can be phase locked to a narrow linewidth laser by creating an optical PLL where the phase error signal is generated from a beat between the narrow linewidth laser and the frequency comb. In the case of the DFG comb, locking $f_{\rm rep}$ will provide a fully stabilized optical comb spectrum. When locked to a narrow linewidth optical reference, e.g. a diode laser locked to a high-finesse cavity, it is possible to simultaneously achieve Hz linewidth for all comb lines. Fig.\,\ref{fig:rf_opt} plots the phase noise of a DFG frequency comb locked to both an RF reference and a narrow line-width optical reference, showing the dramatic noise improvements that are possible when locking to an optical reference. The linewidth of the comb lines were on the Hz level when locked to the optical reference.

A cw laser can be locked to a frequency comb by forming a beat between the laser and the frequency comb. This beat can then be phase locked to an RF reference that determines the offset between the frequency comb line and the cw laser frequency. This establishes an optical phase locked loop (OPLL) between the cw laser and the comb. To achieve a phase lock, sufficient bandwidth is required. Adequate bandwidth is typically given in the case of external cavity diode lasers (ECDL) via current modulation or by implementation of an external AOM.

\section{Characterisation of a frequency comb}
\label{sec:5}

\begin{figure}
\includegraphics[width=\textwidth]{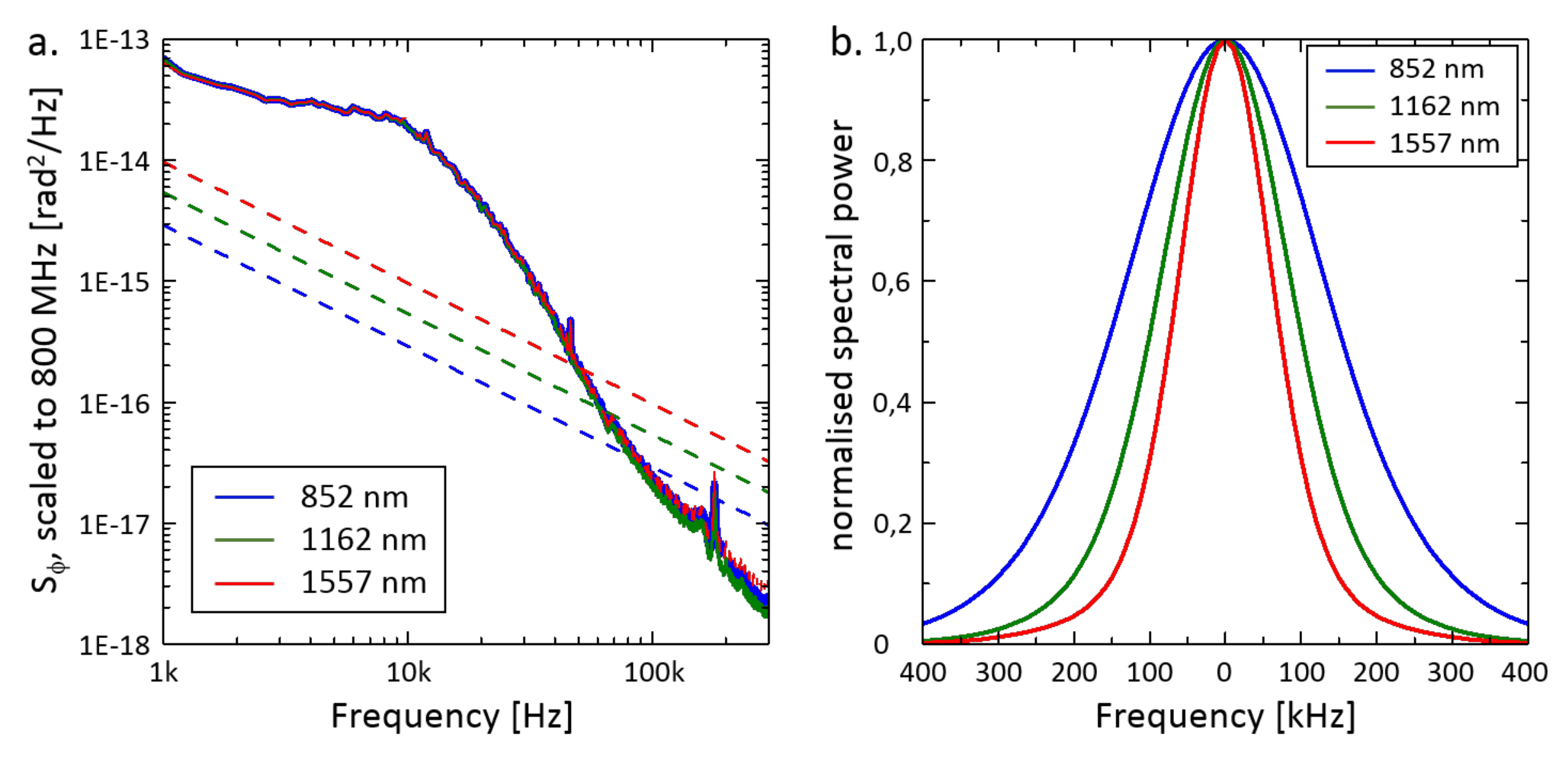}
\caption{Phase noise spectra and reconstructed line shapes of comb teeth at different wavelengths measured on a second DFG system with an additional output at \SI{1162}{\nano \meter}. At \SI{1550}{\nano \meter} and \SI{980}{\nano \meter} an external cavity diode laser is optically phase locked to one comb line. The phase noise is subsequently measured with a delayed self-heterodyne setup (DSH). The noise spectrum at \SI{1162}{\nano \meter} is measured by direct beat with an ECDL locked to a high-finesse cavity. The phase noise is limited by the RF reference and can be further improved by locking to an optical reference (cp Fig.\;\ref{fig:rf_opt}).}
\label{fig:noise}       
\end{figure}

The phase noise of individual comb teeth at different wavelengths can be characterized by a beat with a narrow linewidth laser or, alternatively, by transferring the noise properties to an optically phase-locked (OPL) cw laser. The cw laser can subsequently be characterized using a delayed self-heterodyne (DSH) beat note \cite{Okoshi:80}.  Fig.\;\ref{fig:noise} shows phase noise measurements at different wavelengths measured by DSH of an OPL cw clean-up laser (\SI{1557}{\nano \meter} and \SI{852}{\nano \meter}) and a beat with an ECDL locked to a high-finesse cavity at \SI{1162}{\nano \meter}, respectively. In each case the beat signal is recorded for several seconds, filtered and phase tracked to evaluate the phase noise with respect to the carrier. From the phase noise (Fig.\;\ref{fig:noise}a) the lines shapes and widths can be reconstructed as shown in Fig.\;\ref{fig:noise}b. The $\beta$-separation lines provides a simple concept to the relation between noise and line shape \cite{DiDomenico:10}. It provides a good approximation for a large range of experimentally observed noise spectra: The noise in frequency ranges exceeding the $\beta$-separation line contribute to the carrier linewidth, while the noise below the $\beta$-separation line contribute to the pedestal \cite{Bucalovic:12}. 

\begin{figure}
\includegraphics[width=\textwidth]{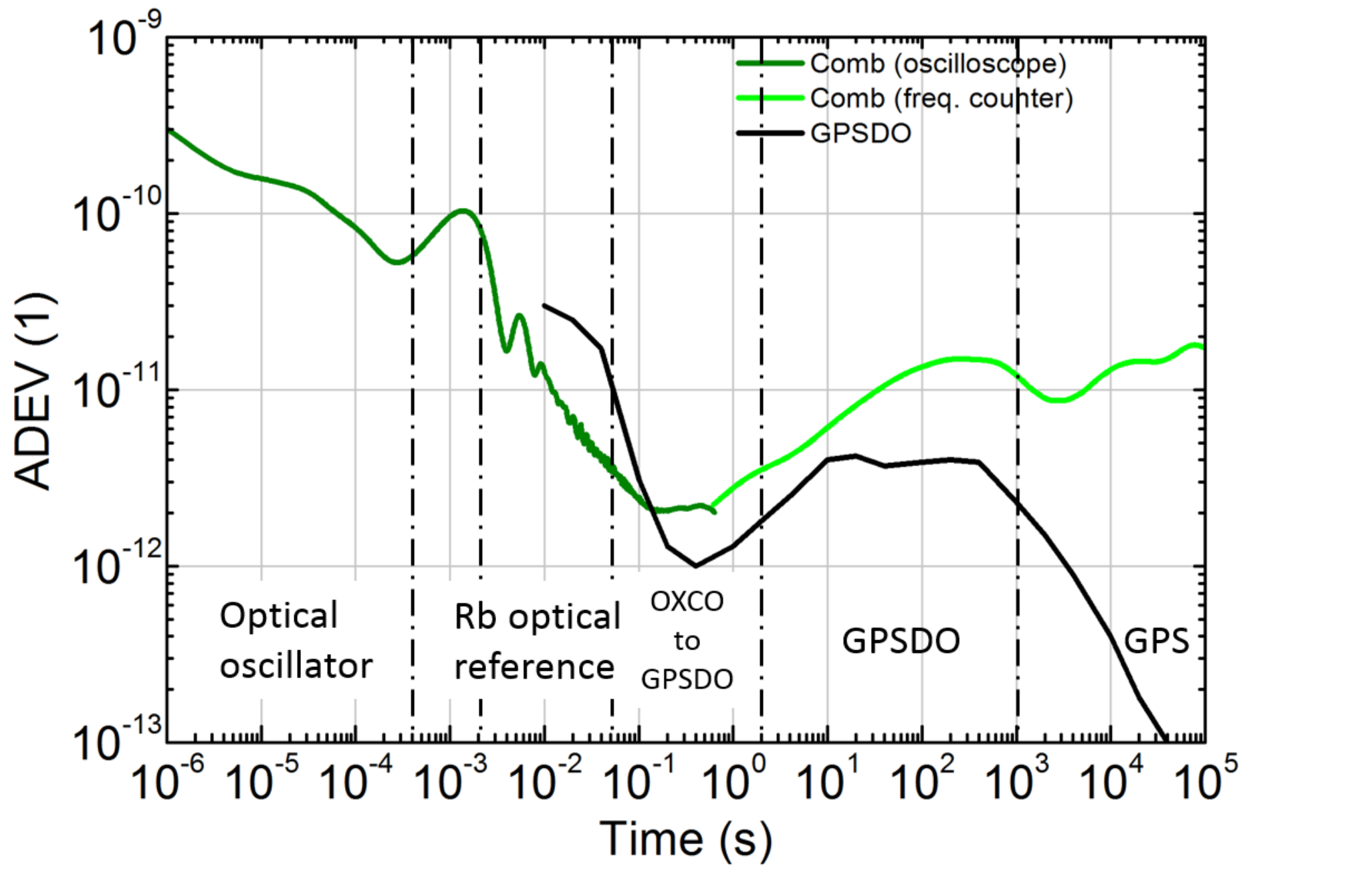}
\caption{Allan deviation of the comb measured against Rb-spectroscopy. The repetition rate $f_{\rm rep}$ is locked to an RF reference at the 10th harmonic at \SI{800}{\mega \hertz}. The Allan deviation of the commercial oven-controlled crystal oscillator based GPS-disciplined RF reference (GPSDO) is shown for reference. The increase of the Allan deviation on long time scales is attributed to drifts in Rb spectroscopy. Note, that an oscillator with lower intrinsic noise shows lower ADEV at short time scales.}
\label{fig:adev}       
\end{figure}

The long term stability of the comb is characterized by the Allan deviation. Fig.\;\ref{fig:adev} shows the long term stability as an Allan deviation (ADEV) measured against Doppler-free saturation spectroscopy of Rb. At short time scales the Allan deviation is extracted from the recorded beat, for longer times it is recorded by a zero-dead-time frequency counter. Different limiting contributions can be identified at different time scales corresponding to the different locking stages. At short time scales the ADEV is limited by the oscillator noise, within the locking bandwidth it is limited by the Rb reference then by the lock to the ultra-low-noise reference oscillator, which in turn is locked to \SI{10}{\mega \hertz} reference finally by the GPS. In this case the limit to the GPS cannot be observed because the local reference given by the Rb spectroscopy drifts, thus limiting the measured ADEV at long time scales.

\section{An example application: precision spectroscopy of Sr Rydberg states}
\label{sec:6}

\begin{figure}
\includegraphics[width=\textwidth]{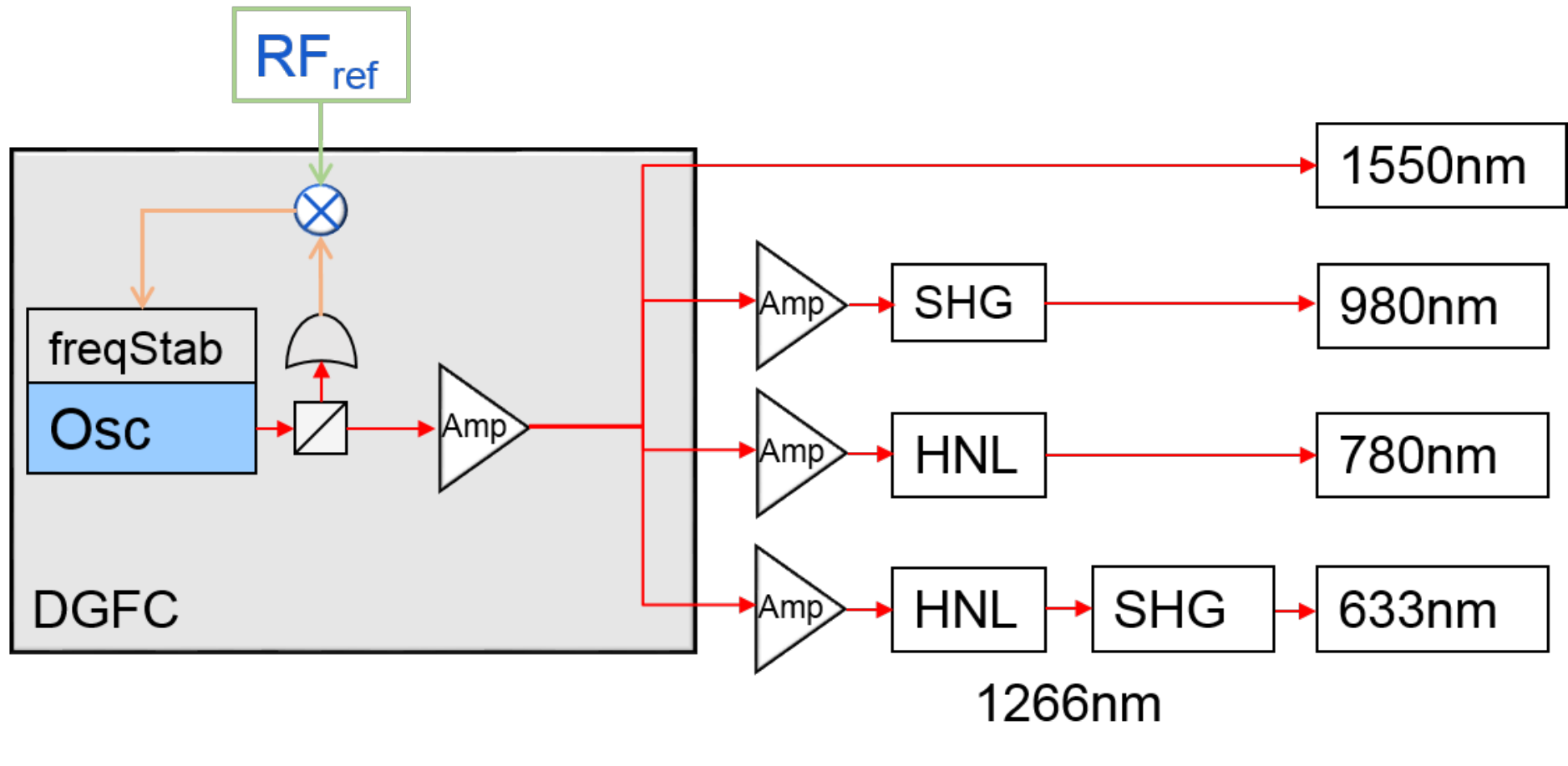}
\caption{Frequency comb setup: The difference frequency comb (DFG-FC) delivers an $f_{\rm ceo}$-free spectrum that can be locked to an RF reference provideing an absolutely stable frequency comb. The stable frequency comb can subsequently be converted to the desired wavelength outputs.}
\label{fig:setup}       
\end{figure}

\begin{figure}
\includegraphics[width=\textwidth]{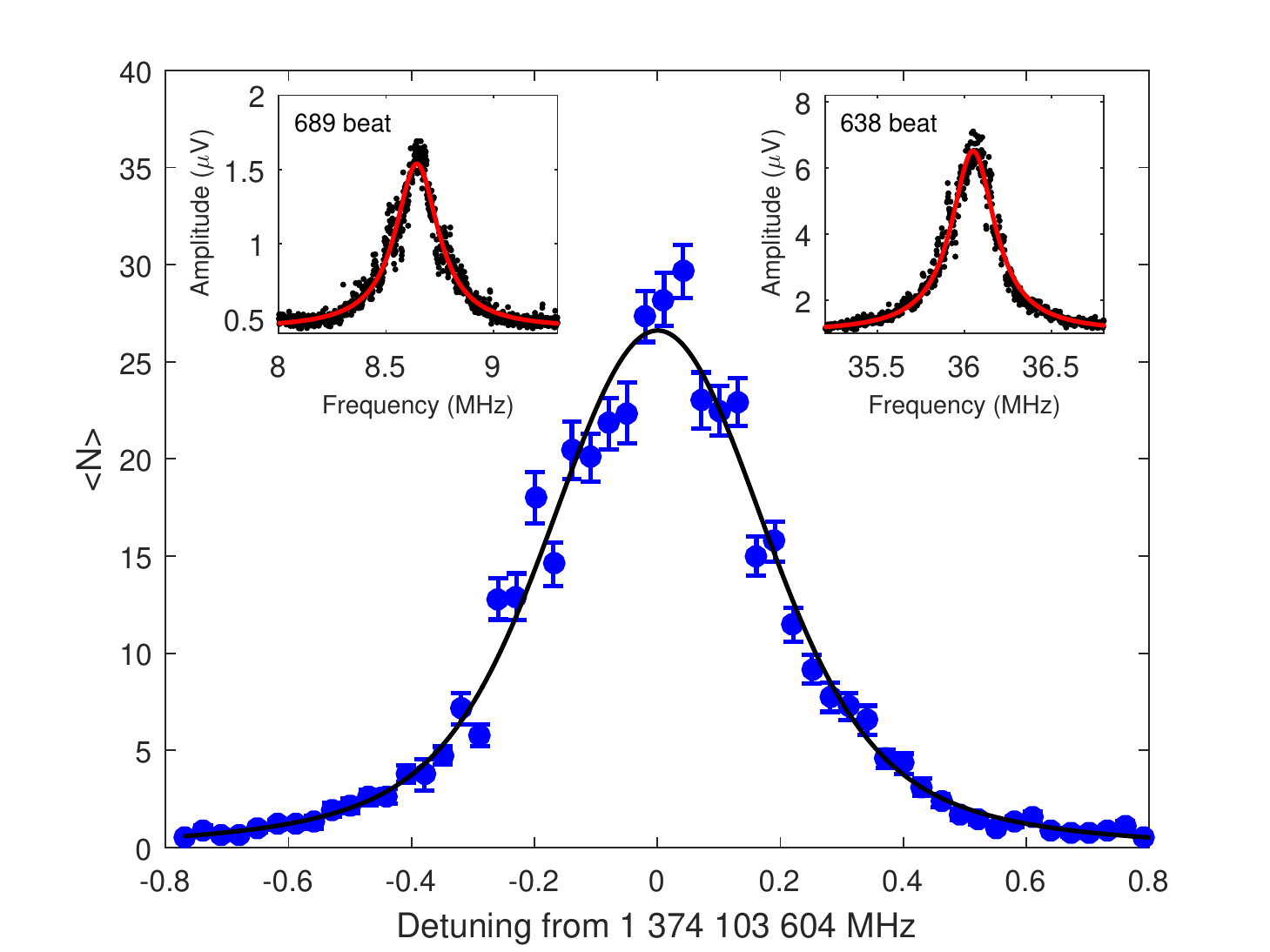}
\caption{Precision Rydberg spectroscopy with the frequency comb. Insets show beat signals between the comb and the CW excitation lasers at \SI{689}{\nano \meter} and \SI{638}{\nano \meter}. Rydberg atoms are excited, ionized and detected in a cold cloud of Sr atoms as described in \cite{Bridge:16}. Using the measured beat frequencies, we obtain a plot of the ion signal $<N>$ versus the absolute frequency of the ($^{88}\mathrm{Sr}\;^1\mathrm{S}_{0} \rightarrow 5s37s\;^3\mathrm{S}_{1}, m_J\!=\!-1$) transition in a \SI{3.1}{\gauss} magnetic field (blue points). Black curve indicates a Gaussian fit to the spectrum.}
\label{fig:Ryd}       
\end{figure}

Precision spectroscopy of Rydberg states can be significantly improved by measuring against a frequency comb.  In strontium, one possible excitation scheme for triplet Rydberg states requires \SI{689}{\nano \meter} and a tunable source at \SI{319}{\nano \meter} \cite{Bridge:16,DeSalvo:16}, these wavelengths have been used to access Rydberg states with principal quantum numbers $n=\!24-\!81$ \cite{Bridge:16,DeSalvo:16,DeSalvo:15}. In \cite{Bridge:16} the UV light is generated by second harmonic generation (SHG) of \SI{638}{\nano \meter}, hence precision Sr Rydberg spectroscopy requires frequency comb outputs at both \SI{638}{\nano \meter} and \SI{689}{\nano \meter} (cp. Fig.\;\ref{fig:setup}).   For the absolute measurement of the frequency, a beat between each laser and a nearby comb line is detected with a photodiode. An example of the measured beat notes, and a Rydberg excitation spectrum with the resulting absolute frequency axis are shown in Fig.\;\ref{fig:Ryd}. Using the least squares fit the absolute frequency of the line center is determined with a statistical uncertainty of just \SI{4}{\kilo \hertz}, which is 3 orders of magnitude better than previous measurements \cite{Beigang:82}. Systematic frequency shifts due to magnetic and electric fields and the blackbody environment affect the line position, and the next challenge is to control these at a similar level. For the data shown in Fig.\;\ref{fig:Ryd}, our preliminary estimates suggest that these are less than \SI{1}{\mega \hertz}.

\section{Conclusion}
\label{sec:7}

Frequency combs have great potential as general purpose tools for high precision measurements of optical frequencies as well as timing applications. In recent years significant scientific and technological advancements have been made both in the noise properties as well as in terms of versatility and integration. Applications require control of both parameters ($f_{\rm ceo}, f_{\rm rep}$) of the comb spectrum. In DFG combs the carrier envelope phase is fundamentally eliminated, removing the limitations of a feed-back-loop stabilization. Moreover, this opens the current of oscillator pump as a fast actuator for other purposes. This allows for a compact and robust, passively stable frequency comb solution. The increasing availability of frequency combs as laboratory tools open this potential to a wide range of precision measurements, in particular in cold atom physics.

\section{Acknowledgement}
NH and RK acknowledge support by Initial Training Networks QTea and COHERENCE, funded by the FP7 Marie Curie Actions of the European Commission ITN-317485 and ITN-265031, respectively. Durham University acknowledge support from EPSRC
grant EP/J007021/ and EU grants FP7-ICT-2013-612862-HAIRS and H2020-FETPROACT-2014-640378-RYSQ.

Some research data supporting this paper is available at {\it link added later}; the remainder is commercially relevant but may be obtained under agreement by contacting the authors.

\bibliography{ITNarxiv}

\end{document}